\documentclass[twocolumn,superscriptaddress,floatfix,preprintnumbers,prd ,nofootinbib,hyperref]{revtex4-2}
\pdfoutput=1
\usepackage[colorlinks=true,breaklinks=true]{hyperref}
\usepackage[normalem]{ulem}
\usepackage{amsmath,amssymb} 
\usepackage{graphicx} 
\usepackage[export]{adjustbox} 
\usepackage{url} 
\usepackage[dvipsnames]{xcolor} 
\usepackage{siunitx} 

\usepackage{float}
\usepackage{microtype}

\usepackage{hyperref} 
\usepackage[capitalize]{cleveref} 
\newcommand\myshade{80} 
\colorlet{mylinkcolor}{ForestGreen}
\colorlet{mycitecolor}{Red}
\colorlet{myurlcolor}{violet}
\hypersetup{
  linkcolor  = mylinkcolor!\myshade!black,
  citecolor  = mycitecolor!\myshade!black,
  urlcolor   = myurlcolor!\myshade!black,
  colorlinks = true
}

\makeatletter

\renewcommand{\p@section}{}

\renewcommand{\p@subsection}{}

\renewcommand{\p@subsubsection}{}
\makeatother

\newcommand{\En}{E_\text{nuc}}

\newcommand{\EM}{\text{EM }}

\newcommand{\iso}[2]{{\ensuremath{{}^{#2}}\ensuremath{\rm #1}}}

\newcommand{\github}[1]{\href{https://github.com/#1}{\includegraphics[width=8pt]{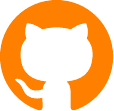}}}

\newcommand{\orcid}[1]{\begingroup
  \hypersetup{hidelinks}\href{https://orcid.org/#1}{\includegraphics[width=10pt]{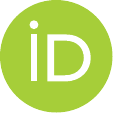}} \endgroup}

\makeatletter
\providecommand*{\diff}%
	{\@ifnextchar^{\DIfF}{\DIfF^{}}}
\def\DIfF^#1{%
	\mathop{\mathrm{\mathstrut d}}%
		\nolimits^{#1}\gobblespace}
\def\gobblespace{%
	\futurelet\diffarg\opspace}
\def\opspace{%
	\let\DiffSpace\!%
	\ifx\diffarg(%
		\let\DiffSpace\relax
	\else
		\ifx\diffarg[%
			\let\DiffSpace\relax
		\else
  			\ifx\diffarg\{%
				\let\DiffSpace\relax
			\fi\fi\fi\DiffSpace}

\usepackage{verbatim}

\begin{document}

\title{On the sensitivity of nuclear clocks to new physics}

\author{Andrea Caputo \orcid{0000-0003-1122-6606}}
\email{andrea.caputo@cern.ch}
\affiliation{Theoretical Physics Department, CERN,
             1211 Geneva 23, Switzerland}

\author{Doron Gazit \orcid{https://orcid.org/0000-0002-0350-3266}}
\email{doron.gazit@mail.huji.ac.il}
\affiliation{Racah Institute of Physics, The Hebrew University,
             9190401 Jerusalem, Israel}

\author{Hans--Werner Hammer \orcid{0000-0002-2318-0644}}
\email{Hans-Werner.Hammer@physik.tu-darmstadt.de}
\affiliation{Technische Universit\"at Darmstadt, Department of Physics,
64289 Darmstadt, Germany}
\affiliation{Helmholtz Forschungsakademie Hessen für FAIR (HFHF) and Extreme Matter Institute EMMI, GSI Helmholtzzentrum für Schwerionenforschung GmbH, 64291 Darmstadt, Germany}

\author{Joachim Kopp \orcid{0000-0003-0600-4996}}
\email{jkopp@cern.ch}
\affiliation{Theoretical Physics Department, CERN,
             1211 Geneva 23, Switzerland}
\affiliation{PRISMA Cluster of Excellence \& Mainz Institute for
             Theoretical Physics, \\
             Johannes Gutenberg University, 55099 Mainz, Germany}

\author{Gil Paz \orcid{0000-0001-5605-2043}}
\email{gilpaz@wayne.edu}
\affiliation{Department of Physics and Astronomy, Wayne State University,
             Detroit, Michigan 48201, USA}

\author{Gilad Perez \orcid{0000-0002-3878-1821}}
\email{gilad.perez@weizmann.ac.il}
\affiliation{Department of Particle Physics and Astrophysics,
             Weizmann Institute of Science, Rehovot 761001, Israel}

\author{Konstantin Springmann \orcid{0000-0002-9617-6136}}
\email{konstantin.springmann@weizmann.ac.il}
\affiliation{Department of Particle Physics and Astrophysics,
             Weizmann Institute of Science, Rehovot 761001, Israel}
\affiliation{Physik-Department, Technische Universit\"at M\"unchen,
             85748 Garching, Germany}

\date{\today}

\preprint{CERN-TH-2024-117, MITP-24-063, TUM-HEP-1515/24, WSU-HEP-2403} 


\begin{abstract}
\noindent
The recent demonstration of laser excitation of the $\approx 8$ eV isomeric state of $\iso{Th}{229\text{m}}$ is a significant step towards a nuclear clock. 
The low excitation energy likely results from a cancellation between electromagnetic and strong contributions, which new physics can disrupt. 
In this Letter, we quantify the enhancement of a nuclear clock’s sensitivity to new physics using a geometric model and a novel $d$-wave halo model of the nucleus that reproduces measured differences between $\iso{Th}{229\text{m}}$ states. 
We find likely enhancements of order $10^4$ while a worst case scenario with enhancement $\ll 1$ is unlikely.
\end{abstract}

\maketitle

\textit{Introduction.} 
The common lore in theoretical physics today is that roughly 80\% of the Universe's matter is dark matter (DM). Despite ample astrophysical and cosmological evidence\footnote{Notice that in our work we assume dark matter, thus new massive particles, to be the most likely explanation for a wide set of observables, from the CMB anisotropies, to galactic rotation curves, passing for lensing probes. However, alternatives to dark matter, such as modified theories of gravity, are also being considered.}, its nature remains a unknown. Ultralight DM (ULDM) theories, including axion~\cite{Preskill:1982cy,Abbott:1982af,Dine:1982ah,Hook:2018dlk,DiLuzio:2020wdo}, dilaton~\cite{Arvanitaki:2014faa,Hubisz:2024hyz}, relaxion~\cite{Banerjee:2018xmn,Chatrchyan:2022dpy}, and Higgs-portal models~\cite{Piazza:2010ye}, offer simple explanations. The Nelson--Barr framework also proposes a viable ULDM candidate~\cite{Dine:2024bxv}. These models suggest ULDM couples mainly to the Standard Model (SM) QCD sector, causing nuclear parameter oscillations~\cite{Flambaum:2006ip,Damour:2010rm,Damour:2010rp}, and subdominantly to the QED sector, affecting $\alpha_\text{em}$. 

Scalar ULDM couples linearly to hadron masses~\cite{Antypas:2022asj}, whereas pseudo-scalars (axions) couple quadratically~\cite{Kim:2022ype,Hook:2017psm,Balkin:2022qer}. There are also natural ULDM models with quadratic DM interaction with SM fields~\cite{Banerjee:2022sqg}, leading to nuclear parameter oscillations.

Searching for variations in SM parameters involves comparing two clocks with different parameter dependencies~\cite{Damour:2010rp,Arvanitaki:2014faa,Stadnik:2015kia}. Laboratory limits come from clock-comparison experiments using atomic/molecular spectroscopy, cavities, and mechanical oscillators. Hyperfine transitions and mechanical oscillators sensitive to nuclear parameters~\cite{Hees:2016gop,Kennedy:2020bac,Campbell:2020fvq,Kobayashi:2022vsf,Zhang:2022ewz,Sherrill:2023zah} lack the accuracy of optical clocks. Optical clocks involve nuclear properties via hyperfine structure and reduced mass, contributing $\sim 10^{-6}$ and $\sim 10^{-5}$, respectively, with charge radius oscillation contributing $\sim 10^{-3}$\cite{Banerjee:2023bjc}. Molecular rovibrational transitions promise $\mathcal{O}(1)$ sensitivity to nuclear parameter modulation\cite{Kozyryev:2018pcp,Madge:2024aot}, though current constraints exist above \SI{10}{\Hz}~\cite{Oswald:2021vtc}. See e.g.\ Ref.~\cite{Safronova:2017xyt} for a review about the use of atoms and molecules for new physics searches.

A \emph{nuclear clock}, monitoring an ultra-narrow \emph{nuclear} transition, could significantly advance ULDM searches~\cite{Peik:2003, Campbell:2012, Kazakov:2012, vonderWense:2020bbs}. 
A suitable isotope is \iso{Th}{229}, with an isomeric state \iso{Th}{229\text{m}} at \SI{8}{eV} above the ground state, excitable with a laser~\cite{Kraemer:2022gpi,Thirolf:2024}. Recent laser excitation observations mark progress towards a nuclear clock~\cite{Tiedau:2024obk,Elwell:2024qyh,Zhang:2024ngu}. The \iso{Th}{229} data's line-shape analysis already shows high sensitivity to physics beyond the SM~\cite{BSMToAppear}.

A nuclear clock's sensitivity to new physics comes from an accidental cancellation between nuclear and electromagnetic (\EM) energy shifts. If disrupted by new physics this can lead to high sensitivity to beyond SM physics, $\sim 10^5$ times~\cite{Flambaum:2006ak, Fadeev:2020bgt, Rellergert:2010zz, Safronova:2019lex, Peik:2020cwm, Hayes, Banks:2024sli}. For recent reviews see~\cite{Antypas:2022asj,Maheshwari:2024cbh,Thirolf:2024}.

Accurately quantifying this sensitivity requires understanding the degree of fine-tuning in the transition energy. All current models of atomic nuclei have intrinsic imperfections, and a worst case scenario with no cancellation, where \EM and nuclear energy differences are individually at the eV scale, cannot be ruled out, leaving the nuclear clock with an enhancement factor of $\mathcal{O}(1)$.

Our goal is to quantify how likely the worst case scenario is. We do so by calculating the Coulomb energy shift between the ground state and the isomer of \iso{Th}{229} first in a classical, geometric, model of the nucleus, parameterized by its charge radius, quadrupole and higher moments, as well as the skin thickness. We then derive similar results using a quantum halo model, in which \iso{Th}{229} is described as a \iso{Th}{228} core orbited by a halo neutron, and the two states differ in the spin alignment of the halo neutron. Finally, we conclude and propose a number of auxiliary measurements and further steps that could shed more light on the likelihood of the worst case scenario and the sensitivity of a nuclear clock to physics beyond the SM.

\textit{The $\iso{ Th}{229}$ Nucleus.} $^{229}$Th is a heavy nucleus, making its modeling from first principles challenging. The isomeric state's energy difference with the ground state, $\Delta E = \SI{8.355733554021(8)}{eV}$~\cite{Zhang:2024ngu}, is much lower than typical nuclear physics scales. This low scale is thought to result from a cancellation between the \EM, $\Delta E_\text{em}$, and nuclear, $\Delta \En$, contributions to the transition energy:
\begin{align}
    \Delta E=\Delta E_\text{em}+\Delta \En \,,
    \label{master}
\end{align}
with $\big|\Delta E_\text{em}\big|\sim \big |\Delta \En\big|\gg \Delta E \,.$ Physics beyond the SM could disrupt this fine-tuned cancellation.

The sensitivity enhancement of a nuclear clock compared to an optical lattice clock is $\big|\Delta E_\text{em} \big|/ \Delta E \simeq \big|\Delta \En \big|/ \Delta E$. To compute this enhancement, it is necessary to determine $\Delta E_\text{em}$ independently. Since $\Delta E_\text{em}$ is not observable and cannot be computed from first principles, modeling is required.

If new physics affects the strong coupling constant, $\delta \alpha_s$, and/or the \EM coupling constant, $\delta \alpha_\text{em}$, we can express the transition energy variation as
\begin{align}
    \delta(\Delta E) = \frac{\partial\Delta E_\text{em}}{\partial \alpha_\text{em}} \delta\alpha_\text{em}
             + \frac{\partial(\Delta E_\text{em} + \Delta \En)}{\partial\alpha_s} \delta\alpha_s,
    \label{Eq:Sens}
\end{align}
where $\partial(\Delta E_\text{em} + \Delta \En) / \partial \alpha_\text{em} = \partial\Delta E_\text{em} / \partial \alpha_\text{em}$. DM-induced variations of $\alpha_{\text{em,s}}$ are typically $\delta \alpha_{\text{em},\text{s}} = \alpha_{\text{em},\text{s}} (1 + \mathcal{F}(t) g_{\text{em},\text{s}})\,$, with $g_{\text{em},\text{s}}$ denoting the DM coupling strength and $\mathcal{F}(t)$ depending on DM oscillations.

We focus on computing the \EM enhancement factor,
\begin{align}
    K_\text{em} \equiv \frac{1}{\Delta E}
                       \frac{\partial\Delta E_\text{em}}{\partial \log \alpha_\text{em}}
                \simeq \frac{\Delta E_\text{em}}{\Delta E}\,,
\end{align}
\textbf{where we used that, to leading order}, $\Delta E_\text{em}$ depends linearly on $\alpha_\text{em}$ in the second equality. 
$K_\text{em}$ is the quantity of interest for DM candidates coupled to the \EM sector; however, such candidates are typically well constrained by other probes.
But some of the most appealing DM candidates couple to the strong sector, and hence it is of great interest to compute the corresponding enhancement factor $K_s=K_s^\text{em}+K_s^\text{nuc}$ related to variations of the effective value of $\alpha_s$.
The effect of such variations is very difficult to model. Nevertheless,
if $\Delta E_{\text{em}}$ has a polynomial dependence on $\alpha_s$ (e.g. from variation of nuclear radius with $\alpha_s$, see~\cite{Banerjee:2023bjc}), we expect
\begin{align}
    K_s^{\text{em}} \equiv \frac{1}{\Delta E} \frac{\partial \Delta E_\text{em}}{\partial\log\alpha_s}
                \sim   \beta \frac{\Delta E_\text{em}}{\Delta E}
                \sim   \beta K_\text{em}\,,
\end{align}
with $\beta$ expected to be order one. This implies $K_\text{em}$ can be a good proxy for $K_s$.
Barring cancellations between $K_{s}^{\text{em}}$ and $K_s^{\text{nuc}} \equiv \frac{1}{\Delta E}\frac{\partial \Delta \En}{\partial \log\alpha_s}$, we expect our computation of $K_\text{em}$ to be a good proxy for $K_s$, and our conclusions to apply also to DM models which couple to the strong sector. This statement can be invalidated only if the functional dependence of $\Delta E_\text{em}$ and $\Delta \En$ on $\alpha_s$ is exactly the same.

We now determine $\Delta E_\text{em}$ and its variation with $\alpha_\text{em}$, defining the enhancement factor as  $K\equiv\left\lvert K_\text{em}\right\rvert$. Recent studies modeled the \iso{Th}{229} nucleus using a geometric model \cite{Berengut:2009zz, Fadeev:2020bgt}, describing the nucleus as a three-dimensional charged body. This model, though simple, has shortcomings, including neglecting quantum effects and exchange terms \cite{Naito:2020jrc}. An alternative halo model, describing \iso{Th}{229} as a \iso{Th}{228} core with a halo neutron, is supported by neutron separation energy data and charge radii comparisons \cite{Hammer:2017tjm, Browne:2008otd, Abusaleem:2014bsy}. We investigate the predictions of both models quantitatively. We  neglect any contribution from the electrons density change inside the nucleus. Following Ref.~\cite{Dzuba:2023muh}, the electrons contribution to the energy shift is negligible, $\Delta E = F \delta \langle r^2 \rangle \simeq 10^{-7} \, \text{eV}$ (where $F$ is the field shift, $\sim \rm GHz/fm^2$~\cite{Dzuba:2023muh}).

\begin{figure*}[ht]
\vskip-5pt
\includegraphics[width=0.90\columnwidth]{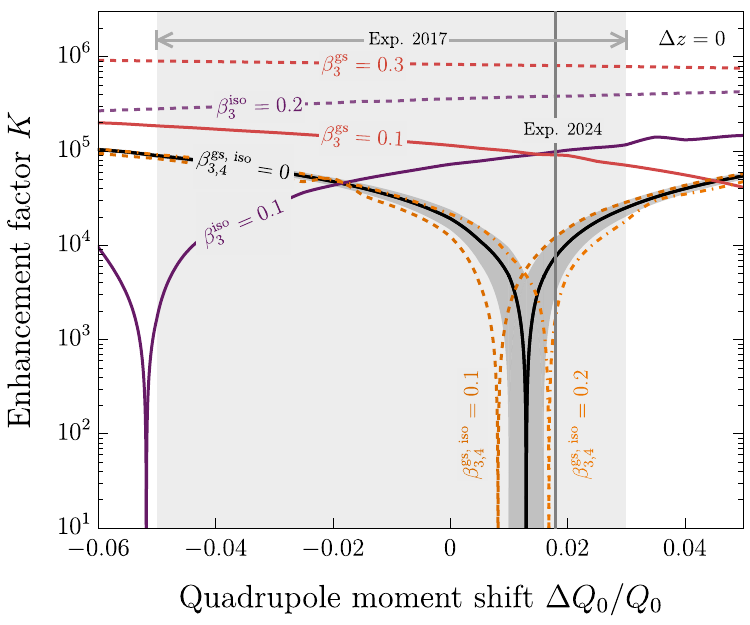}
\hspace{30pt}
\includegraphics[width=0.90\columnwidth]{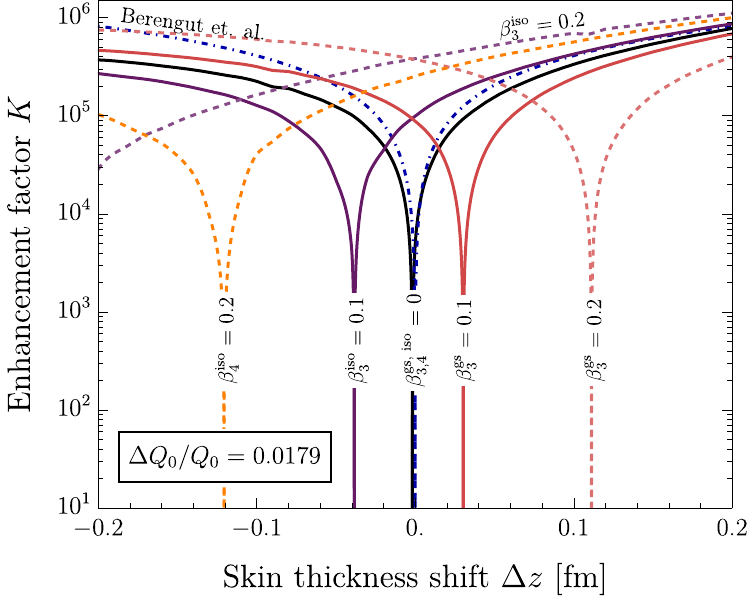}
\vskip-5pt
\caption{\textbf{Left panel.} The enhancement factor $K$ benefiting new physics searches with \iso{Th}{229} as a function of the quadrupole moment difference between the ground and isomer states. Colored lines correspond to different values of the higher \EM multipole moments $\beta_{3,4}^\text{gs,iso}$ of the ground state (``gs'') and the isomer (``iso''), cf.\ \cref{eq:R-theta}. Labels give the multipole moments which are chosen different from zero. The vertical gray line indicates the latest, very precise, measured value of $\Delta Q_0/Q_0$~\cite{Zhang:2024ngu}, while the wide shaded gray band corresponds to the $1\sigma$ uncertainty from previous measurements of the quadrupole moments~\cite{Thielking:2017qet}. The dark-gray shaded area denotes the $2\sigma$ uncertainty from $\langle r^2 \rangle$ for the case of $\beta_{3,4}^{\text{gs, iso}}=0$. We see that for specific values of $\Delta Q_0/Q_0$ and the values of $\beta_{3,4}^{\text{gs, iso}}$, the $K$-factor may be drastically smaller than  $10^5$. \textbf{Right panel.} $K$ as a function of $\Delta z$, the difference in the skin thickness between the ground and isomer states of \iso{Th}{229}. Also in this case we see that $K$ can go to zero for specific values of $\Delta z$ and $\beta_{3,4}^{\text{gs, iso}}$, though for most values of these unknown parameters, $K$ is of order $10^4$--$10^5$. The dot-dashed blue line shows the approximate expression for $K$ from Eq.~7 of Ref.~\cite{Berengut:2009zz}, which does not take into account the impact of $\beta_{3,4}$ and is a good approximation only for small $\Delta z$~\cite{DEVRIES1987495}.}
\label{fig:AllGeometrical}
\vskip-10pt
\end{figure*}

\textit{The classical geometric model.} 
Despite its aforementioned shortcomings, the geometric model of the \iso{Th}{229} nucleus has been adopted in several studies of nuclear clocks' potential to look for new physics~\cite{Berengut:2009zz, Fadeev:2020bgt}. We therefore study this model first, advancing it beyond previous calculations. Calculating $\Delta E_\text{em}$ at the required level of $\sim \SI{8}{eV}$ is considered absurd in nuclear physics, and we do not claim to achieve this either. Instead, we aim to estimate a preferred range of $\Delta E_\text{em}$, and answer the following two questions:

\begin{enumerate}
    \item What is the expected enhancement factor $K$?
    \item How likely is the worst case scenario, where $\Delta E_\text{em}$ is accidentally close to zero, jeopardising new physics searches?
\end{enumerate}

Our starting point is the nuclear charge density, which we describe with a Woods--Saxon distribution~\cite{PhysRev.95.577}
\begin{align} \label{eq:Woods-Saxon}
    \rho(r,\theta) = \rho_0\left[1 + \exp\left(\frac{r - R(\theta)}{z}\right)\right]^{-1},
\end{align}
where $z$ is the ``skin thickness" of the nucleus, 
$R$ is its ``radius" and the constant $\rho_0$ is determined by the normalisation condition $Z \, e = \int d^3\mathbf{r} \,\rho(r,\theta)$ with $e$ the electric charge unit, and $Z=90$ for thorium. We model non-sphericity of the nucleus through $R(\theta)$:
\begin{align}
   \frac{R(\theta)}{R_0}=  1 + \beta_2 Y_{20}(\theta)
                            + \beta_3 Y_{30}(\theta)
                            + \beta_4 Y_{40}(\theta) +\dots\, ,
    \label{eq:R-theta}
\end{align}
where $R_0$ is the scale radius, $Y_{lm}(\theta,\phi)$ denote spherical harmonics, $\beta_2$, $\beta_3$, and $\beta_4$ are coefficients of the quadrupole, octupole, and hexadecapole, respectively. Higher multipole moments are neglected in this work. Azimuthal symmetry sets the magnetic quantum number $m$ to zero, and $R(\theta)$ depends only on the polar angle $\theta$. Note that $\beta_3$ and $\beta_4$ are not constrained by experiments yet, contrary to $\beta_2$, but they all impact $\Delta E_{\text{em}}$.

Given $\rho(r,\theta)$, we compute its two moments which are accessible experimentally: the mean squared charge radius
\begin{align}\label{eq:r2}
    \langle r^2 \rangle \equiv \frac{1}{e \, Z}\,\int d^3 \mathbf{r}\,r^2 \rho(r,\theta) \, ,
\end{align}
and the intrinsic quadrupole moment\footnote{We use the notation $Q_0$ here to distinguish the intrinsic quadrupole moment from the spectroscopic quadrupole moment $Q_s = Q_0 [ (j (2 j - 1)]/[(j + 1) (2 j + 3)]$, which is often the quantity that is reported in the experimental literature.}
\begin{align}
\label{eq:quadrupole}
    Q_0 \equiv \frac{1}{eZ}\int d^3 \mathbf{r}\,r^2 \rho(r,\theta)\left[3 \cos^2(\theta) - 1\right] \, .
\end{align}
The measurements of $\langle r^2 \rangle$ and $Q_0$ fix two parameters of our model. In particular, the scale radius $R_0$ depends mainly on $\langle r^2 \rangle = (\SI{5.7557(143)}{fm})^2$ \cite{Safronova:2018fue, Thielking:2017qet}, while the parameter $\beta_2$ is closely related to the intrinsic quadrupole moment, $Q_0 = \SI{9.8(1)}{fm^2}$~\cite{Thielking:2017qet}. From this value $\beta_2\sim 0.2$ with weak dependence on $\beta_{3,4}$.
For small $\beta_{2,3,4}$ one finds, $\langle r^2\rangle\sim R_0^2$ plus small $\mathcal{O}( z^2/R_0^2)$ corrections (typically $z\sim0.5\,\text{fm}$).
Similarly, for small $\beta_2$, $Q_0$ can be expanded in $\beta_2$. In the following, we do not assume any expansions and numerically solve for various parameters of our model.

Note that Ref.~\cite{Fadeev:2020bgt} relates $\langle r^2 \rangle$ to $\beta_2$ (or equivalently $Q_0$) by taking a constant volume approach, that is, by assuming equal $R_0$ for the ground state and the isomer.
Furthermore, \cite{Fadeev:2020bgt} studies the impact of small $\beta_3$ values on the energy difference $\Delta E_{\text{em}}$, but not the impact of $\beta_4$.
Our study does not assume constant volume. We numerically scan over the parameters $\beta_{3,4}\in\left[0,0.3\right]$, i.e. of the order of $\beta_2$, and $\Delta z$, the difference in the skin thickness between the ground and isomer states of \iso{Th}{229}, to carefully asses their combined impact on the sensitivity.

To calculate the \EM energy for both the ground state and the isomer of \iso{Th}{229}, we approximate them with the direct Coulomb energy contribution
\begin{align}
    E_\text{em}\simeq E_{\rm C}
    \equiv \frac{1}{2} \int \! d^3\mathbf{r} \, d^3\mathbf{r}' \,
              \frac{\rho(r,\theta) \, \rho(r',\theta')}{\left| \mathbf{r} - \mathbf{r}' \right|}.
    \label{Eq:CoulombEnergy}
\end{align}
The integral is calculated numerically.  We show the results in \cref{fig:AllGeometrical}.

In the left panel of \cref{fig:AllGeometrical} we plot the enhancement factor $K$ as a function of the relative quadrupole moment difference between the isomer state and the ground state of \iso{Th}{229}, $\Delta Q_0 / Q_0$, for various combinations of $\beta_{3,4}$.  We find that the worst case scenario in which $K \simeq 0$, implying no enhanced sensitivity to new physics, is fairly unlikely. It would only be realized for finely-tuned values of $\beta_{3,4}^\text{gs,iso}$ of the ground state (``gs'') and the isomer (``iso'') such that the funnel region is aligned with the measured value of $\Delta Q_0 / Q_0$ from Ref.~\cite{Zhang:2024ngu} (vertical dark gray line in left panel of \cref{fig:AllGeometrical}), but typical values of $K \sim 10^4$--$10^5$ are far more likely. 
Same conclusions can be drawn from Fig. S1 in the supplementary material: K factor as a function of $\beta_{3,4}^{\text{gs, iso}}$. Similar fine-tuning is needed for a given value of  $\langle r^2 \rangle$. We illustrate this by the dark-gray shaded area denoting the $2\sigma$ uncertainty from $\langle r^2 \rangle$~\cite{Safronova:2018fue}. For a precisely measured value of $\langle r^2 \rangle$, $\beta_{3,4}^{\text{gs, iso}}$ must be fine-tuned to zero to obtain small values of $K$.

In the right panel of \cref{fig:AllGeometrical} we show the impact of $\Delta z$ for both the ground state and the isomer. We see again that $K$ might be very small or zero. Note that the treatment of $\Delta z$ discussed in \cite{Berengut:2009zz} is valid only for small $\Delta z$ and does not simultaneously take into account the effects of $\beta_{3,4}$.

\textit{Th-229 as a Halo Nucleus.} To describe \iso{Th}{229} as a halo nucleus, we follow the halo Effective Field Theory (EFT) framework for shallow $d$-wave states developed in \cite{Braun:2018vez, Braun:2018hug}. We consider the ground state ($j^\pi = 5/2^+$) and the isomer ($j^\pi = 3/2^+$) as a doublet resulting from the $l=2$ coupling of the halo neutron to the $\iso{Th}{228}$ core system. We stress that halo EFT should only be considered a model for the neutron--\iso{Th}{228} system, not a proper EFT, as the separation between the low and high energy scales is marginal. If the expansion parameter is estimated as the ratio of the neutron separation energies of \iso{Th}{229} ($S_n \approx \SI{5.2}{MeV}$) and \iso{Th}{228} ($S_n \approx \SI{7.1}{MeV}$), it would be $\sqrt{5.2/7.1} \approx 0.86$. In addition, the predictivity of halo EFT is limited for $d$-wave nuclei as the centrifugal barrier creates a strong dependence on short-distance (small $r$) effects and therefore on the overlap of the halo nucleon's wave function with the core. This is reflected in a leading-order dependence of most observables on counter-terms whose exact magnitude depends on the core--halo interaction model.

However, $d$-wave halo EFT suggests that the neutron-halo doublet states, $j={3/2}$ and $j={5/2}$, are degenerate at leading order, and that therefore the \emph{differences} between their charge radii, quadrupole moments, etc., vanish at leading order \cite{Braun:2018hug,Braun:2018vez}. Thus, we develop here a model, inspired by halo EFT, that focuses on these differences, motivated by the nature of the isomeric transition. 
We begin by computing the two quantities to which we have experimental access: $\langle r^2 \rangle$ and $Q_0$. 
Within our halo model, the charge density of the ground and isomer states are identical, except for a difference from the spin--orbit interaction between the halo neutron's spin and its $l=2$ orbital angular momentum. This spin--orbit contribution to the charge density is given by~\cite{Ong:2010gf}
\begin{align}
\label{eq:ChargeDensitySO}
    \rho_\text{SO} = \frac{\mu_n}{2 m_n^2}  i \sigma_{\alpha \beta} 
                     \Big[ (\nabla \Phi_{\alpha}^{\dagger}) \times (\nabla \Phi_{\beta}) \Big]\,,
\end{align}
where $\mu_n = -1.91$ is the magnetic moment of the neutron in units of the nuclear magneton, $\Phi_\alpha$ is the (non-relativistic 2-component) neutron spinor with $\alpha, \beta = 1,2$ spinor indices. With $\rho_\text{SO}$ at hand we can compute the corresponding mean squared charge radius and the quadrupole moment, in analogy to our calculations for the geometric model, using Eq.~\ref{eq:r2} and Eq.~\ref{eq:quadrupole}. The calculations appear in the supplementary material \cite{supp}.

The difference of the mean squared radii 
\begin{align}
    \Delta \langle r^2 \rangle_\text{SO}
        \equiv \langle r^2 \rangle_\text{SO}^\text{iso} - \langle r^2 \rangle_\text{SO}^\text{gs}
        =      \SI{0.0047}{fm^2}\,,
\end{align}
is smaller than the measured $\Delta \langle r^2 \rangle = \SI{0.012(2)}{fm^2}$~\cite{Thielking:2017qet} by a factor of $\sim 2$, which we consider excellent agreement given the inherent uncertainties to the halo model. For the quadrupole moment, we find
\begin{align}
    Q_\text{SO}^\text{iso} - Q_\text{SO}^\text{gs} = \SI{0.185}{fm^2}\,,
\end{align}
in astonishingly good agreement with the latest experimental value~\cite{Zhang:2024ngu}
\begin{align}
    (Q^\text{iso} - Q^\text{gs})_\text{exp}
        = \SI{0.173(9)}{fm^2}.
\end{align}
We consider this a further validation of the halo model for \iso{Th}{229}. We also notice that the computations of $\Delta \langle r^2 \rangle_\text{SO}$ and $Q_\text{SO}^\text{iso} - Q_\text{SO}^\text{gs}$ are \textit{not} sensitive to ultraviolet (UV) physics, i.e., insensitive to the structure of the halo neutron wave function near the origin. This is not expected to be true for higher multiple moments.

The main assumption in these calculations is that the contribution originates from single nucleons~\cite{Ong:2010gf}, i.e., consistent with a halo model~\cite{Hammer:2017tjm}. This leaves room for a possible renormalization of the halo-neutron magnetic moment, which offers an explanation for the $\approx 50\%$ deviation from experiment.

Having shown that the halo model gives sensible results for $\langle r^2 \rangle$ and $Q_0$, we compute the spin--orbit contribution to the binding energy of the halo neutron~\cite{Sakurai:QM}
\begin{equation}
    \begin{aligned} \label{eq:E_SO}
    E_\text{SO} = \frac{e\,\mu_n}{2 m_n^2}
                  \bigg[ j (j + 1) - l (l + 1) - \frac{3}{4} \bigg]
                 \int_{0}^{\infty} \left[ \frac{[u(r)]^2}{r}
                                           \frac{dV_{\rm C}}{dr} \right] dr \,,
\end{aligned}
\vspace{0.5cm}
\end{equation}
where $l=2$ and $j = 5/2$ (ground state) or $3/2$ (isomer) are its angular momentum quantum numbers, $V_{\rm C}(r)$ is the Coulomb potential of the \iso{Th}{228} core, and $u(r)$ is the excess neutron's radial $d$-wave function~\cite{VolyaZelevinsky:NP}: 
\begin{align}
    u(r) = A(r) e^{-\gamma r} \Big( 1 + \frac{3}{\gamma r}
    + \frac{3}{(\gamma r)^2} \Big)\,.
    \label{eq:halowave}
\end{align}
In this expression, $\gamma \equiv \sqrt{2 m_n E_B}$, $E_B = \SI{5.2}{MeV}$ is the binding energy, and  $A(r)$ is determined by short-range physics.  $A(r)$ has a weak dependence on $r$, and it asymptotically converges at large $r$ to a constant $A$, called the asymptotic normalization constant (ANC). At leading order in halo EFT, $A$ is taken to be fully independent of $r$, and depends on the UV cutoff scale, which we assume in the following. As usual in EFTs, the behavior of the ANC at short distances should be fixed from additional experimental constraints (see the discussion in the outlook). Since the neutron separation energy for the ground state of \iso{Th}{229} is expected to be almost the same as for the isomer, a good assumption is that $u(r)$ is also the same for $j = 5/2$ and $j =3/2$. After all, these quantum numbers are just a result of the spin--orbit coupling for fixed $l=2$, and we treat the spin--orbit potential as a small perturbation.

In order to estimate $E_\text{SO}$ from \cref{eq:E_SO}, we also need to specify the Coulomb potential, $V_{\rm C}(r)$. As the halo-neutron radial wave function is dominated by the length scale $\gamma^{-1}$, i.e., a few Fermi and much larger than the UV length-scale, the charge density can be approximated by a Woods--Saxon form \cite{CottinghamGreenwood:NP},  see \cref{eq:Woods-Saxon}. The universal property $\rho \approx \text{constant}$ leads to $dV_{\rm C} / dr \propto r$, which conveniently cancels the UV sensitivity of the integral in \cref{eq:E_SO}; we then assume a constant $A(r)$, constrained by the normalization condition on $u(r)$, and numerically find
\begin{align}
    \Delta E_\text{SO} = \left.E_\text{SO}\right\rvert_{j=5/2} - E_\text{SO}|_{j=3/2}
                       \approx 144\, p \, \text{keV}\,,
\end{align}
where the order-one factor $p$ is a function of $A$ and $l$, the  latter being a short distance cut-off we introduced to regularize the integrals for $r \rightarrow 0$ (we find that the integral is almost independent of $l$ for values $l \leq \SI{4}{fm}$). A more complete nuclear model would predict a specific form for $A(r)$, eliminating the need for a regulator, and thereby determining $p$. Intuitively, $p$ is related to the probability of finding the halo neutron inside the nucleus. The details of the computation are given in the supplementary material~\cite{supp}.

Summarizing, we find the contribution of the spin--orbit interaction to $\Delta E_\text{em}$ to be of order \SI{100}{keV}, which corresponds to an enhancement factor $K \approx 10^4$ for new physics searches using nuclear clocks.

\textit{Summary and Outlook.} In this work, we studied how the sensitivity of a \iso{Th}{229} nuclear clock to new physics depends on nuclear modeling. We showed that within a classical geometric model of the \iso{Th}{229} nucleus, a ``worst case scenario" where the sensitivity goes to zero is possible but requires fine-tuning of the model parameters at the per cent or per mille level. Conversely, we presented a more realistic quantum halo model, where an unpaired neutron of \iso{Th}{229} is loosely bound to a \iso{Th}{228} spin-0 core in a $d$-wave state with orbital angular momentum $l=2$. We validated this model against experimentally measured differences in charge radii and quadrupole moments of the \iso{Th}{229} ground state and isomer \iso{Th}{229\text{m}}, correctly predicting these differences to be close to zero. Within the halo model, the ``worst case scenario" is never realized, and the enhancement factor $K_\text{em}$ for new physics searches is always $\gtrsim 10^4$.

Our work suggests several future research directions. The main obstacle to accurately determining the enhancement factor is the nuclear model, with its many unknown parameters. Future neutron capture experiments could refine and verify the halo model. The amplitude of neutron capture in $d$-wave halos is proportional to the asymptotic normalization constant $A$, so such experiments can constrain the probability of finding the halo-neutron inside the core, affecting the spin-orbit contribution to the \EM interaction. We assumed $A$ is identical for the ground state and isomer, an assumption that can be tested by measuring the neutron capture cross section, $\sigma_\text{cap}$, for both states. Halo EFT predicts $\sigma_\text{cap}^\text{iso} = \frac{2}{3} \sigma_\text{cap}^\text{gs}$ \cite{Braun:2018vez}.

It is important to confirm the latest results for $\Delta Q_0/Q_0$~\cite{Zhang:2024ngu} and improve the determination of $\Delta\langle r^2 \rangle$. Additionally, devising tests sensitive to higher-order multipoles $\beta_{3,4}$ and skin thickness $z$ is highly desirable. These quantities are crucial for determining the enhancement factor within the classical geometric model, which, despite its shortcomings, is still of interest and represents an alternative to the $d$-wave halo model.

Finally, we argued that $K_\text{em} \sim K_{s}$, so our results apply to new physics coupled to either the \EM or strong sectors. This holds unless the \EM and nuclear energy differences between the ground and isomer states of thorium-229 have the exact same dependence on $\alpha_s$. EFT approaches to the expansion of nuclear forces in the QCD scale predict different dependencies on this scale~\cite{Meissner:2014pma,Hammer:2019poc}. Verifying this within concrete frameworks is key. We leave this and other developments for future work.

\bibliographystyle{bibi}
\bibliography{biblio}

\end{document}